\documentclass[conference]{IEEEtran}

\usepackage{cite}

\usepackage{graphicx}
\usepackage{xcolor}
\usepackage{tikz}
\usetikzlibrary{arrows,positioning,shapes.geometric}
\usepackage{pgfplots}
\usepackage{multirow}
\usepackage{upgreek}
\usepackage{amssymb}
\usepackage{amsmath}
\usepackage{cases}
\usepackage{pifont}
\usepackage{bm}
\usepackage{amsthm}

\usepackage{tabularx}
\usepackage{booktabs}
\usepackage{filecontents}
\usepackage{subcaption}
\newcolumntype{Y}{>{\centering\arraybackslash}X}

\usepackage[noend,ruled]{algorithm2e}

\usepackage{array}

\newcommand{\fixme}[2]{\ifx&#2&{\leavevmode\color{red}#1}\else{\leavevmode\color{red}FIXME\{}#1{\leavevmode\color{red}\}}\footnote{{\leavevmode\color{red}#2}}\PackageWarning{Fixme}{#1: #2}\fi}

\DeclareMathOperator{\PM}{PM}

\title{Blind Detection with Polar Codes}

\author{Carlo~Condo, Seyyed~Ali~Hashemi, Warren~J.~Gross}
\thanks{C.~Condo, S.~A.~Hashemi and W.~J.~Gross are with the Department of Electrical and Computer Engineering, McGill University, Montr\'eal, Qu\'ebec, Canada. e-mail: carlo.condo@mail.mcgill.ca, seyyed.hashemi@mail.mcgill.ca, warren.gross@mcgill.ca.}

\begin{document}

\maketitle

\begin{abstract}
In blind detection, a set of candidates has to be decoded within a strict time constraint, to identify which transmissions are directed at the user equipment. Blind detection is an operation required by the 3GPP LTE/LTE-Advanced standard, and it will be required in the $5^{\text{th}}$ generation wireless communication standard (5G) as well.
We propose a blind detection scheme based on polar codes, where the radio network temporary identifier (RNTI) is transmitted instead of some of the frozen bits. A low-complexity decoding stage decodes all candidates, selecting a subset that is decoded by a high-performance algorithm. Simulations results show good missed detection and false alarm rates, that meet the system specifications. We also propose an early stopping criterion for the second decoding stage that can reduce the number of operations performed, improving both average latency and energy consumption. The detection speed is analyzed and different system parameter combinations are shown to meet the stringent timing requirements, leading to various implementation trade-offs.
\end{abstract}

\section{Introduction} \label{sec:intro}

Blind decoding, or blind detection, is an operation foreseen by the 3GPP LTE/LTE-Advanced standards to allow the user equipment (UE) to gather control information related to the downlink shared channel. The UE attempts the decoding of a set of candidates determined by combinations of system parameters, to identify if one of the candidates holds its control information. The scheme used in LTE relies on the concatenation of a cyclic redundancy check (CRC) with a convolutional code.

Blind detection will be present also in the $5^{\text{th}}$ generation wireless communication standard (5G): ongoing discussions are considering a substantial reduction of the time frame allocated to blind detection, from $16\mu$s to $4\mu$s. Blind detection must be performed very frequently, and given the high number of decoding attempts that must be performed in a limited time \cite{3GPP_R8}, it can lead to large implementation costs and high energy consumption.

Polar codes are linear block codes, with proven capacity-achieving property and a low-complexity encoding and decoding process \cite{arikan}. They have been chosen to be adopted in 5G \cite{3gpp_polar}. Successive-cancellation (SC) is the first polar code decoding algorithm: while optimal for infinite code lengths, it grants mediocre error-correction performance at moderate and short code lengths. In its standard formulation, it also has long decoding latency. SC list (SCL) decoding has been proposed in \cite{tal_list} to improve the error-correction performance of SC, sacrificing speed. Subsequent works \cite{sarkis, xiong_symbol, hashemi_SSCL, hashemi_FSSCL} have proposed improvements to both SC and SCL decoding speed.

The blind detection of polar codes has been independently researched in the recent work \cite{Giard_BD}, where a detection metric based on constituent codes has been developed.
In this paper, we propose a blind detection scheme based on polar codes. A first SC decoding stage helps selecting a set of candidates, subsequently decoded with SCL. The scheme is evaluated in terms of error-correction capability, missed detections and false alarms, showing its compliance with the requirements of the standard. An early stopping criterion for SCL is also proposed to reduce energy consumption and average latency. The detection speed is analyzed, identifying possible combinations of system parameters to meet the standard current and future timing constraints.


\section{Preliminaries} \label{sec:prel}
\subsection{Polar Codes}

A polar code of length $N=2^n$ and rate $K/N$, denoted as $\mathcal{P}(N,K)$, is a linear block code that can be expressed as the concatenation of two polar codes of length $N/2$. This recursive construction is represented by a modulo-$2$ matrix multiplication as $\mathbf{x} = \mathbf{u} \mathbf{G}^{\otimes n}$,
where $\mathbf{u} = \{u_0,u_1,\ldots,u_{N-1}\}$ is the input vector, $\mathbf{x} = \{x_0,x_1,\ldots,x_{N-1}\}$ is the codeword, and the generator matrix $\mathbf{G}^{\otimes n}$ is the $n$-th Kronecker product of the polarizing matrix $\mathbf{G}=\bigl[\begin{smallmatrix} 1&0\\ 1&1 \end{smallmatrix} \bigr]$. The polar code structure allows to identify, in the $N$-bit input vector $\mathbf{u}$, reliable and unreliable bit-channels.
The $K$ information bits are assigned to the most reliable bit-channels of $\mathbf{u}$, while the remaining $N-K$, called frozen bits, are set to a predefined value, usually $0$.
Codeword $\mathbf{x}$ is transmitted through the channel, and the decoder receives the Logarithmic Likelihood Ratio (LLR) vector $\mathbf{y} = \{y_0,y_1,\ldots,y_{N-1}\}$.

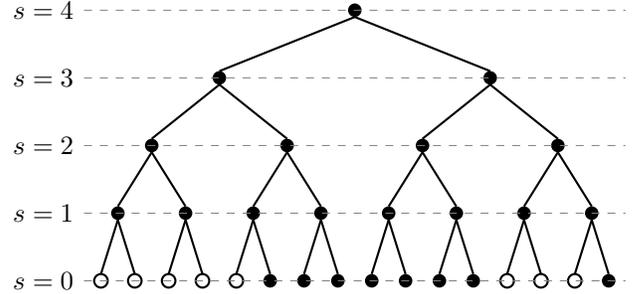
\begin{figure}
\centering
\begin{tikzpicture}[scale=1.8, thick]

\fill (0,0) circle [radius=.05];

\fill (-1,-.5) circle [radius=.05];
\fill (1,-.5) circle [radius=.05];

\fill (-1.5,-1) circle [radius=.05];
\fill  (-.5,-1) circle [radius=.05];
\fill  (.5,-1) circle [radius=.05];
\fill  (1.5,-1) circle [radius=.05];

\fill  (-1.75,-1.5) circle [radius=.05];
\fill  (-1.25,-1.5) circle [radius=.05];
\fill (-.75,-1.5) circle [radius=.05];
\fill (-.25,-1.5) circle [radius=.05];
\fill  (.25,-1.5) circle [radius=.05];
\fill  (.75,-1.5) circle [radius=.05];
\fill  (1.25,-1.5) circle [radius=.05];
\fill (1.75,-1.5) circle [radius=.05];

\draw (-1.875,-2) circle [radius=.05];
\draw  (-1.625,-2) circle [radius=.05];
\draw  (-1.375,-2) circle [radius=.05];
\draw  (-1.125,-2) circle [radius=.05];
\draw  (-.875,-2) circle [radius=.05];
\fill  (-.625,-2) circle [radius=.05];
\fill  (-.375,-2) circle [radius=.05];
\fill  (-.125,-2) circle [radius=.05];
\fill  (.125,-2) circle [radius=.05];
\fill  (.375,-2) circle [radius=.05];
\fill  (.625,-2) circle [radius=.05];
\fill  (.875,-2) circle [radius=.05];
\draw  (1.125,-2) circle [radius=.05];
\draw  (1.375,-2) circle [radius=.05];
\draw  (1.625,-2) circle [radius=.05];
\fill  (1.875,-2) circle [radius=.05];

\draw (0,-.05) -- (-1,-.45);
\draw (0,-.05) -- (1,-.45);

\draw (-1,-.55) -- (-1.5,-.95);
\draw (-1,-.55) -- (-.5,-.95);
\draw (1,-.55) -- (.5,-.95);
\draw (1,-.55) -- (1.5,-.95);

\draw  (-1.5,-1.05) -- (-1.75,-1.45);
\draw  (-1.5,-1.05) -- (-1.25,-1.45);
\draw (-.5,-1.05) -- (-.75,-1.45);
\draw (-.5,-1.05) -- (-.25,-1.45);
\draw  (.5,-1.05) -- (.25,-1.45);
\draw  (.5,-1.05) -- (.75,-1.45);
\draw  (1.5,-1.05) -- (1.25,-1.45);
\draw  (1.5,-1.05) -- (1.75,-1.45);

\draw  (-1.75,-1.55) -- (-1.875,-1.95);
\draw  (-1.75,-1.55) -- (-1.625,-1.95);
\draw  (-1.25,-1.55) -- (-1.375,-1.95);
\draw  (-1.25,-1.55) -- (-1.125,-1.95);
\draw  (-.75,-1.55) -- (-.875,-1.95);
\draw  (-.75,-1.55) -- (-.625,-1.95);
\draw (-.25,-1.55) -- (-.375,-1.95);
\draw (-.25,-1.55) -- (-.125,-1.95);
\draw  (.25,-1.55) -- (.125,-1.95);
\draw (.25,-1.55) -- (.375,-1.95);
\draw (.75,-1.55) -- (.625,-1.95);
\draw  (.75,-1.55) -- (.875,-1.95);
\draw  (1.25,-1.55) -- (1.125,-1.95);
\draw  (1.25,-1.55) -- (1.375,-1.95);
\draw  (1.75,-1.55) -- (1.625,-1.95);
\draw (1.75,-1.55) -- (1.875,-1.95);

\draw [very thin,gray,dashed] (-2,0) -- (2,0);
\draw [very thin,gray,dashed] (-2,-.5) -- (2,-.5);
\draw [very thin,gray,dashed] (-2,-1) -- (2,-1);
\draw [very thin,gray,dashed] (-2,-1.5) -- (2,-1.5);
\draw [very thin,gray,dashed] (-2,-2) -- (2,-2);

\node at (-2.3,0) {$s=4$};
\node at (-2.3,-.5) {$s=3$};
\node at (-2.3,-1) {$s=2$};
\node at (-2.3,-1.5) {$s=1$};
\node at (-2.3,-2) {$s=0$};

\end{tikzpicture}
\caption{Binary tree example for $\mathcal{P}(16,8)$. White circles at $s=0$ are frozen bits, black circles at $s=0$ are information bits.}
\label{fig:tree}
\end{figure}

Along with the definition of polar codes, in \cite{arikan}, the SC decoder is proposed. The SC-based decoding process can be represented as a binary tree search, in which the tree is explored depth first, with priority to the left branches. Fig.~\ref{fig:tree} shows an example of SC decoding tree for $\mathcal{P}(16,8)$, where nodes at stage $s$ contain $2^s$ bits. White leaf nodes are frozen bits, while black leaf nodes are information bits.

\begin{figure}
\centering
\begin{tikzpicture}[scale=.5]

\draw [very thin,gray,dashed] (-2,0) -- (2,0);
\draw [very thin,gray,dashed] (-2,-2) -- (2,-2);
\draw [very thin,gray,dashed] (-2,-4) -- (2,-4);

\node at (-3,0) {$s+1$};
\node at (-3,-2) {$s$};
\node at (-3,-4) {$s-1$};

\fill (0,0) circle [radius=.25];
\fill  (0,-2) circle [radius=.2];
\fill  (-1.5,-4) circle [radius=.15];
\fill  (1.5,-4) circle [radius=.15];

\draw [->,very thick] (-.1,-.4) -- (-.1,-1.7) node [left,midway,rotate=0] {$\alpha$};
\draw [->,very thick] (.1,-1.7) -- (.1,-.4) node [right,midway,rotate=0] {$\beta$};

\draw [->,very thick] (-.25,-2.2) -- (-1.45,-3.75) node [left,midway,rotate=0] {$\alpha^{\text{l}}$};
\draw [->,very thick] (-1.3,-3.85) -- (-.1,-2.3) node [right,near start,rotate=0] {$\beta^{\text{l}}$};
\draw [<-,very thick] (.25,-2.2) -- (1.45,-3.75) node [right,midway,rotate=0] {$\beta^{\text{r}}$};
\draw [<-,very thick] (1.3,-3.85) -- (.1,-2.3) node [left,near start,rotate=0] {$\alpha^{\text{r}}$};

\end{tikzpicture}
\caption{Message passing in tree graph representation of SC decoding.}
\label{fig:MessagePassing}
\end{figure}
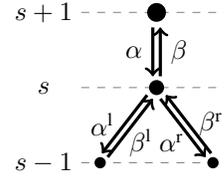

Fig.~\ref{fig:MessagePassing} portrays the message passing among SC tree nodes. Parents pass LLR values $\alpha$ to children, that send in return the hard bit estimates $\beta$. 
The left and right branch messages $\alpha^\text{l}$ and $\alpha^\text{r}$, in the hardware-friendly version of \cite{leroux}, are computed as
\begin{align}
\alpha^{\text{l}}_i = & \text{sgn}(\alpha_i)\text{sgn}(\alpha_{i+2^{s-1}})\min(|\alpha_i|,|\alpha_{i+2^{s-1}}|) \\
\alpha^{\text{r}}_i =& \alpha_{i+2^{s-1}} + (1-2\beta^\text{l}_i)\alpha_i \text{,}
\label{eq2}
\end{align}
while $\beta$ is computed as
\begin{equation}
\beta_i =
  \begin{cases}
    \beta^\text{l}_i\oplus \beta^\text{r}_i, & \text{if} \quad i < 2^{s-1}\\
    \beta^\text{r}_{i-2^{s-1}}, & \text{otherwise},
  \end{cases}
  \label{eq3}
\end{equation}
where $\oplus$ denotes the bitwise XOR. The SC operations are scheduled according to the following order: each node receives $\alpha$ first, then sends $\alpha^\text{l}$, receives $\beta^\text{l}$, sends $\alpha^\text{r}$, receives $\beta^\text{r}$, and finally sends $\beta$.
When a leaf node is reached, $\beta_i$ is set as the estimated bit $\hat{u}_i$:
\begin{equation}
\hat{u}_i =
  \begin{cases}
    0 \text{,} & \text{if } i \in \mathcal{F} \text{ or } \alpha_{i}\geq 0\text{,}\\
    1 \text{,} & \text{otherwise,}
  \end{cases} \label{eq6}
\end{equation}
where $\mathcal{F}$ is the set of frozen bits.

The SC decoding process requires full tree exploration: however, in \cite{alamdar, sarkis} it has been shown that it is possible to prune the tree by identifying patterns in the sequence of frozen and information bits, achieving substantial speed increments. This improved SC decoding is called fast simplified SC (Fast-SSC).


SC decoding suffers from modest error correction performance with moderate and short code lengths. To improve it, the SCL algorithm was proposed in \cite{tal_list}. It is based on the same process as SC, but each time that a bit is estimated at a leaf node, both its possible values $0$ and $1$ are considered. A set of $L$ codeword candidates is stored, so that a bit estimation results in $2L$ new candidates, half of which must be discarded. To this purpose, a Path Metric (PM) is associated to each candidate and updated at every new estimate: the $L$ paths with the lowest PM survive. In the LLR-based SCL proposed in \cite{balatsoukas}, the hardware-friendly formulation of the PM is
\begin{align}
 \text{PM}_{{i}_l} =& \begin{cases}
    \text{PM}_{{i-1}_l}, & \text{if } \hat{u}_{i_l} = \frac{1}{2}\left(1-\text{sgn}\left(\alpha_{i_l}\right)\right)\text{,}\\
    \text{PM}_{{i-1}_l} + |\alpha_{i_l}|, & \text{otherwise,}
  \end{cases} \label{eq7}
\end{align} 
where $l$ is the path index and $\hat{u}_{j_l}$ is the estimate of bit $j$ at path $l$.
As with SC decoding, SCL tree pruning techniques relying on the identification of frozen-information bit patterns have been proposed in \cite{hashemi_SSCL,hashemi_FSSCL}, called simplified SCL (SSCL) and Fast-SSCL.

\subsection{Blind Detection}

The physical downlink control channel (PDCCH) is used in 3GPP LTE/LTE-Advanced to transmit the downlink control information (DCI) related to the downlink shared channel. The DCI carries information regarding the channel resource allocation, transport format and hybrid automatic repeat request, and allows the UE to receive, demodulate and decode.

A CRC is attached to the DCI payload before transmission. The CRC is masked according to the radio network temporary identifier (RNTI) of the UE to which the transmission is directed, or according to one of the system-wide RNTIs. Finally, the DCI is encoded with a convolutional code. The UE is not aware of the format with which the DCI has been transmitted: it thus has to explore a combination of PDCCH locations, PDCCH formats, and DCI formats in the common search space (CSS) and UE-specific search space (UESSS) and attempt decoding to identify useful DCIs. This process is called blind decoding, or blind detection. For each PDCCH candidate in the search space, the UE performs channel decoding, and demasks the CRC with its UE RNTI. If no error is found in the CRC, the DCI is considered as carrying the UE control information.

Based on LTE standard R8 \cite{3GPP_R8}, the performance specifications for the blind detection process are the following:
\begin{itemize}
 \item The DCI of PDCCH is from $8$ to $57$ bits plus $16$-bit CRC, masked by $16$-bit RNTI.
 \item In UESSS, a maximum of $2$ DCI formats can be sent per transmission time interval (TTI) for $2$ potential frame lengths. Therefore, $16$ candidate locations in UESSS $\rightarrow$ $32$ candidates.
 \item In CSS, a maximum of $2$ DCI formats can be sent per TTI for $2$ potential frame lengths. Therefore, $6$ candidate locations in CSS $\rightarrow$ $12$ candidates.
 \item Code length could be between $72$ and $576$ bits.
 \item Information length (including $16$-bit CRC) could be between $24$ and $73$ bits.
 \item Target signal-to-noise ratio (SNR) is dependent on the targeted block error rate (BLER): $10^{-2}$.
 \item There are two types of false-alarm scenarios: Type-1, when the UE RNTI is not transmitted but detected, and Type-2, when the UE RNTI is transmitted but another one is detected. The target false-alarm rate (FAR) is below $10^{-4}$.
 \item Missed detection occurs when UE RNTI is transmitted but not detected. The missed detection rate (MDR) is close to BLER curve.
 \item The available time frame for blind detection is $16\mu$s.
\end{itemize}

\section{Proposed Blind Detection Scheme} \label{sec:blind}

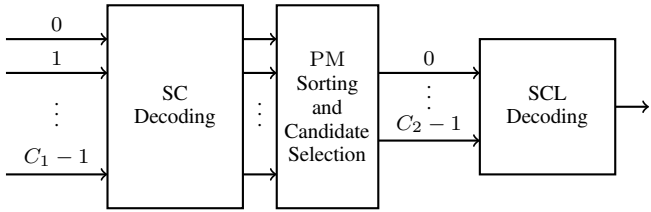
\begin{figure}[t!]
  \centering
  \begin{tikzpicture}[scale=0.9, thick]
\footnotesize
\draw [fill=white] (0,0) rectangle ++(2,3) node [pos=.5,align=center] {SC \\ Decoding};

\draw [->] (-1.5,2.5) -- ++(1.5,0) node [midway, above, sloped] {$0$};
\draw [->] (-1.5,2) -- ++(1.5,0) node [midway, above, sloped] {$1$};

\node at (-.75,1.5) {$\vdots$};

\draw [->] (-1.5,.5) -- ++(1.5,0) node [midway, above, sloped] {$C_1-1$};

\draw [->] (2,2.5) -- ++(.5,0);
\draw [->] (2,2) -- ++(.5,0);

\node at (2.25,1.5) {$\vdots$};

\draw [->] (2,.5) -- ++(.5,0);

\draw [fill=white] (2.5,0) rectangle ++(1.5,3) node [pos=.5,align=center] {$\PM$ \\ Sorting \\ and \\Candidate \\ Selection};

\draw [->] (4,2) -- ++(1.5,0) node [midway, above, sloped] {$0$};

\node at (4.75,1.75) {$\vdots$};

\draw [->] (4,1) -- ++(1.5,0) node [midway, above, sloped] {$C_2-1$};

\draw [fill=white] (5.5,.5) rectangle ++(2,2) node [pos=.5,align=center] {SCL \\ Decoding};

\draw [->] (7.5,1.5) -- ++(.5,0);

\end{tikzpicture}
  \caption{Blind detection with polar codes scheme.}
  \label{fig:scheme}
\end{figure}

We propose the use of polar codes in a blind detection framework, and provide a novel blind detection scheme. In particular, we avoid the use of a CRC, by using some of the frozen bit positions to instead transmit the RNTI. Fig.~\ref{fig:scheme} shows the block diagram of the devised blind detection scheme. $C_1$ candidates are received at the same time: in our case, $C_1=44$. The $C_1$ candidates are decoded with the SC algorithm: the short code lengths considered by the standard allow to keep the latency in check. Moreover, the low implementation complexity of SC allows to have multiple decoders in parallel. A PM is obtained for each candidate: the PM is equivalent to the LLR of the last decoded bit. The PMs are then sorted, to help the selection of the best candidates to forward to the following decoding stage. $C_2$ candidates are in fact selected to be decoded with the powerful SCL decoding algorithm. SCL has a better error correction performance, but a higher implementation complexity. The $C_2$ candidates are chosen as:
\begin{enumerate}
 \item All candidates whose RNTI, after SC decoding, matches the one assigned to the UE. If more than $C_2$ are present, the ones with the highest PMs are selected.
 \item If free slots among the $C_2$ remain, the candidates with the smallest PMs are selected. The candidates with large PMs have higher probability to be correctly decoded: if their RNTI does not match the one assigned to the UE, it is probably a different one. On the other hand, candidates with small PMs have a higher chance of being incorrectly decoded, and a transmission to the UE might be hiding among them.
\end{enumerate}
After SCL decoding, if one of the $C_2$ candidates matches the UE RNTI, it is selected, otherwise no selection is attempted.

\section{Simulation Results} \label{sec:simres}

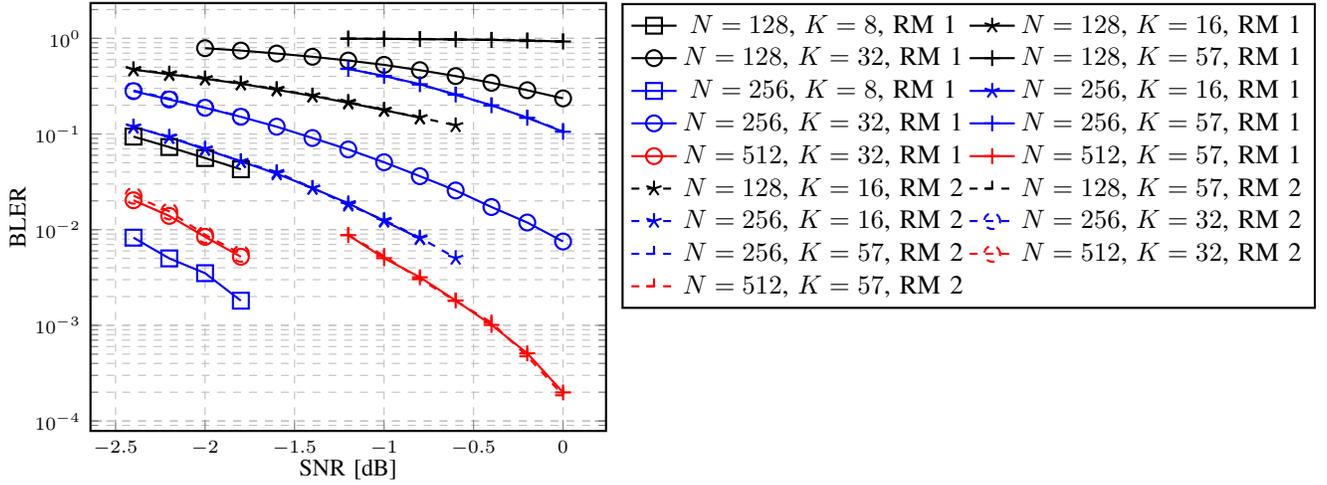
\begin{figure*}[t!]
  \centering
  \begin{tikzpicture}
  \pgfplotsset{
    label style = {font=\fontsize{9pt}{7.2}\selectfont},
    tick label style = {font=\fontsize{7pt}{7.2}\selectfont},
  }

\begin{axis}[
	scale = 1,
    ymode=log,
    xlabel={SNR [dB]}, xlabel style={yshift=0.8em},
    ylabel={BLER}, ylabel style={yshift=-0.75em},%
    grid=both,
    ymajorgrids=true,
    xmajorgrids=true,
    grid style=dashed,
    thick,
    mark size=3,
    legend columns=2,
    legend pos=outer north east,
]

\addplot[
    color=black,
    mark=square,
    thick,
    mark size=3,
]
table {
-2.4 0.09391
-2.2 0.07311
-2.0 0.05611
-1.8 0.04279
};
\addlegendentry{$N=128$, $K=8$, RM 1}

\addplot[
    color=black,
    mark=star,
    thick,
    mark size=3,
]
table {
-2.4 0.47169
-2.2 0.42041
-2.0 0.37742
-1.8 0.33589
-1.6 0.28986
-1.4 0.25062
-1.2 0.21169
-1.0 0.17762
-0.8 0.14792
};
\addlegendentry{$N=128$, $K=16$, RM 1}

\addplot[
    color=black,
    mark=o,
    thick,
    mark size=3,
]
table {
-2.0 0.78901
-1.8 0.74534
-1.6 0.6938
-1.4 0.64212
-1.2 0.58637
-1.0 0.52822
-0.8 0.46485
-0.6 0.40325
-0.4 0.34296
-0.2 0.28678
-0.0 0.23579
};
\addlegendentry{$N=128$, $K=32$, RM 1}

\addplot[
    color=black,
    mark=+,
    thick,
    mark size=3,
]
table {
-1.2 0.99172
-1.0 0.98726
-0.8 0.9814
-0.6 0.97353
-0.4 0.96237
-0.2 0.94687
-0.0 0.92725
};
\addlegendentry{$N=128$, $K=57$, RM 1}

\addplot[
    color=blue,
    mark=square,
    thick,
    mark size=3,
]
table {
-2.4 0.00825
-2.2 0.00501
-2.0 0.00351
-1.8 0.00181
};
\addlegendentry{$N=256$, $K=8$, RM 1}

\addplot[
    color=blue,
    mark=star,
    thick,
    mark size=3,
]
table {
-2.4 0.11941
-2.2 0.09227
-2.0 0.06925
-1.8 0.05172
-1.6 0.03852
-1.4 0.02723
-1.2 0.01886
-1.0 0.01233
-0.8 0.00811
};
\addlegendentry{$N=256$, $K=16$, RM 1}

\addplot[
    color=blue,
    mark=o,
    thick,
    mark size=3,
]
table {
-2.4 0.28056
-2.2 0.22964
-2.0 0.18799
-1.8 0.15138
-1.6 0.1192
-1.4 0.0908
-1.2 0.06899
-1.0 0.05069
-0.8 0.03625
-0.6 0.02569
-0.4 0.01725
-0.2 0.01188
-0.0 0.00752
};
\addlegendentry{$N=256$, $K=32$, RM 1}

\addplot[
    color=blue,
    mark=+,
    thick,
    mark size=3,
]
table {
-1.2 0.47898
-1.0 0.40462
-0.8 0.33153
-0.6 0.25875
-0.4 0.19968
-0.2 0.14778
-0.0 0.10561
};
\addlegendentry{$N=256$, $K=57$, RM 1}

\addplot[
    color=red,
    mark=o,
    thick,
    mark size=3,
]
table {
-2.4 0.02029
-2.2 0.01393
-2.0 0.00842
-1.8 0.00521
};
\addlegendentry{$N=512$, $K=32$, RM 1}

\addplot[
    color=red,
    mark=+,
    thick,
    mark size=3,
]
table {
-1.2 0.00877
-1.0 0.00504
-0.8 0.00318
-0.6 0.00182
-0.4 0.00101
-0.2 0.00050986
-0.0 0.000199016
};
\addlegendentry{$N=512$, $K=57$, RM 1}

\addplot[
    color=black,
    mark=star,
    thick,
    dashed,
    mark size=3,
]
table {
-2.4 0.47406
-2.2 0.42888
-2.0 0.38175
-1.8 0.33754
-1.6 0.29533
-1.4 0.25243
-1.2 0.21659
-1.0 0.17853
-0.8 0.14903
-0.6 0.12299
};
\addlegendentry{$N=128$, $K=16$, RM 2}

\addplot[
    color=black,
    mark=+,
    thick,
    dashed,
    mark size=3,
]
table {
-1.2 0.99223
-1.0 0.98816
-0.8 0.98242
-0.6 0.97439
-0.4 0.96474
-0.2 0.94868 
-0.0 0.92917
};
\addlegendentry{$N=128$, $K=57$, RM 2}

\addplot[
    color=blue,
    mark=star,
    thick,
    dashed,
    mark size=3,
]
table {
-2.4 0.11856
-2.2 0.09358
-2.0 0.07036
-1.8 0.05189
-1.6 0.0399
-1.4 0.02732
-1.2 0.01818
-1.0 0.01259
-0.8 0.00827
-0.6 0.00505
};
\addlegendentry{$N=256$, $K=16$, RM 2}

\addplot[
    color=blue,
    mark=o,
    thick,
    dashed,
    mark size=3,
]
table {
-2.4 0.28454
-2.2 0.23482
-2.0 0.1887
-1.8 0.15248
};
\addlegendentry{$N=256$, $K=32$, RM 2}

\addplot[
    color=blue,
    mark=+,
    thick,
    dashed,
    mark size=3,
]
table {
-1.2 0.48192
-1.0 0.40217
-0.8 0.32964
-0.6 0.25743
-0.4 0.19913
-0.2 0.14829
-0.0 0.10739
};
\addlegendentry{$N=256$, $K=57$, RM 2}

\addplot[
    color=red,
    mark=o,
    thick,
    dashed,
    mark size=3,
]
table {
-2.4 0.02267
-2.2 0.01551
-2.0 0.00884
-1.8 0.0056
};
\addlegendentry{$N=512$, $K=32$, RM 2}

\addplot[
    color=red,
    mark=+,
    thick,
    dashed,
    mark size=3,
]
table {
-1.2 0.00877
-1.0 0.00531
-0.8 0.00304
-0.6 0.0018
-0.4 0.00107
-0.2 0.000474089
-0.0 0.000185668
};
\addlegendentry{$N=512$, $K=57$, RM 2}

\end{axis}
\end{tikzpicture}
  \caption{BLER curves after SC decoding.}
  \label{fig:BLER}
\end{figure*}

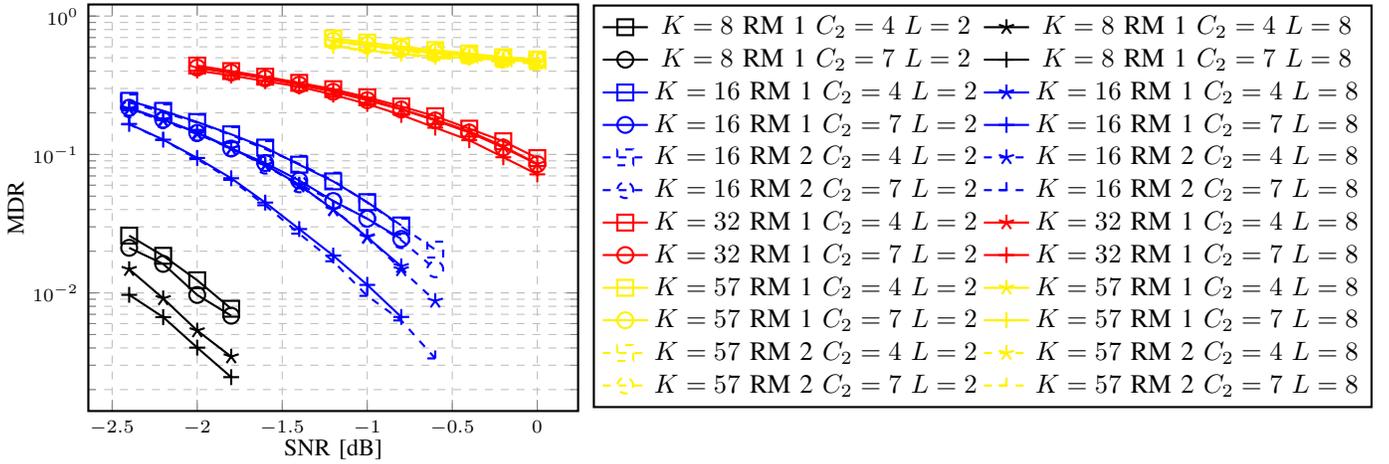
\begin{figure*}[t!]
  \centering
  \begin{tikzpicture}
  \pgfplotsset{
    label style = {font=\fontsize{9pt}{7.2}\selectfont},
    tick label style = {font=\fontsize{7pt}{7.2}\selectfont},
  }

\begin{axis}[
	scale = 0.95,
    ymode=log,
    xlabel={SNR [dB]}, xlabel style={yshift=0.8em},
    ylabel={MDR}, ylabel style={yshift=-0.75em},%
    grid=both,
    ymajorgrids=true,
    xmajorgrids=true,
    grid style=dashed,
    thick,
    mark size=3,
    legend columns=2,
    legend pos=outer north east,
]

\addplot[
    color=black,
    mark=square,
    thick,
    mark size=3,
]
table {
-2.4 0.02592
-2.2 0.01854
-2.0 0.01234
-1.8 0.00768
};
\addlegendentry{$K=8$ RM 1 $C_2=4$ $L=2$}

\addplot[
    color=black,
    mark=star,
    thick,
    mark size=3,
]
table {
-2.4 0.01482
-2.2 0.00912
-2.0 0.00534
-1.8 0.00346
};
\addlegendentry{$K=8$ RM 1 $C_2=4$ $L=8$}

\addplot[
    color=black,
    mark=o,
    thick,
    mark size=3,
]
table {
-2.4 0.02116
-2.2 0.0162
-2.0 0.00966
-1.8 0.00684
};
\addlegendentry{ $K=8$ RM 1 $C_2=7$ $L=2$}

\addplot[
    color=black,
    mark=+,
    thick,
    mark size=3,
]
table {
-2.4 0.00968
-2.2 0.00668
-2.0 0.00402
-1.8 0.00246
};
\addlegendentry{ $K=8$ RM 1 $C_2=7$ $L=8$}

\addplot[
    color=blue,
    mark=square,
    thick,
    mark size=3,
]
table {
-2.4 0.2444
-2.2 0.20686
-2.0 0.17182
-1.8 0.13992
-1.6 0.11222
-1.4 0.08444
-1.2 0.06394
-1.0 0.04526
-0.8 0.03072
};
\addlegendentry{ $K=16$ RM 1 $C_2=4$ $L=2$}

\addplot[
    color=blue,
    mark=star,
    thick,
    mark size=3,
]
table {
-2.4 0.21528
-2.2 0.17852
-2.0 0.14454
-1.8 0.11106
-1.6 0.0841
-1.4 0.06046
-1.2 0.04
-1.0 0.02524
-0.8 0.01556
};
\addlegendentry{ $K=16$ RM 1 $C_2=4$ $L=8$}

\addplot[
    color=blue,
    mark=o,
    thick,
    mark size=3,
]
table {
-2.4 0.21774
-2.2 0.17852
-2.0 0.1434
-1.8 0.11042
-1.6 0.08682
-1.4 0.06492
-1.2 0.04612
-1.0 0.03438
-0.8 0.02454
};
\addlegendentry{ $K=16$ RM 1 $C_2=7$ $L=2$}

\addplot[
    color=blue,
    mark=+,
    thick,
    mark size=3,
]
table {
-2.4 0.16628
-2.2 0.12798
-2.0 0.09438
-1.8 0.06734
-1.6 0.04494
-1.4 0.02892
-1.2 0.0186
-1.0 0.01142
-0.8 0.0067
};
\addlegendentry{ $K=16$ RM 1 $C_2=7$ $L=8$}

\addplot[
    color=blue,
    mark=square,
    thick,
    dashed,
    mark size=3,
]
table {
-2.4 0.2377
-2.2 0.20176
-2.0 0.17014
-1.8 0.13838
-1.6 0.10952
-1.4 0.08562
-1.2 0.06474 
-1.0 0.04496
-0.8 0.03016
-0.6 0.02054
};
\addlegendentry{ $K=16$ RM 2 $C_2=4$ $L=2$}

\addplot[
    color=blue,
    mark=star,
    thick,
    dashed,
    mark size=3,
]
table {
-2.4 0.21256
-2.2 0.1756
-2.0 0.14074
-1.8 0.11046
-1.6 0.08366
-1.4 0.05958
-1.2 0.0403
-1.0 0.02554
-0.8 0.01476
-0.6 0.00874
};
\addlegendentry{ $K=16$ RM 2 $C_2=4$ $L=8$}

\addplot[
    color=blue,
    mark=o,
    thick,
    dashed,
    mark size=3,
]
table {
-2.4 0.21276
-2.2 0.17664
-2.0 0.14248
-1.8 0.11102
-1.6 0.08312
-1.4 0.06086
-1.2 0.04616
-1.0 0.03428
-0.8 0.02404
-0.6 0.01482
};
\addlegendentry{ $K=16$ RM 2 $C_2=7$ $L=2$}

\addplot[
    color=blue,
    mark=+,
    thick,
    dashed,
    mark size=3,
]
table {
-2.4 0.16522
-2.2 0.12742
-2.0 0.09264
-1.8 0.0656
-1.6 0.043
-1.4 0.0268
-1.2 0.01692
-1.0 0.00952
-0.8 0.00632
-0.6 0.00336
};
\addlegendentry{ $K=16$ RM 2 $C_2=7$ $L=8$}

\addplot[
    color=red,
    mark=square,
    thick,
    mark size=3,
]
table {
-2.0 0.43906
-1.8 0.4007
-1.6 0.36506
-1.4 0.32952
-1.2 0.29664
-1.0 0.25916
-0.8 0.22258
-0.6 0.18908
-0.4 0.15374
-0.2 0.12466
-0.0 0.0942
};
\addlegendentry{ $K=32$ RM 1 $C_2=4$ $L=2$}

\addplot[
    color=red,
    mark=star,
    thick,
    mark size=3,
]
table {
-2.0 0.42244
-1.8 0.38912
-1.6 0.35696
-1.4 0.32648
-1.2 0.28848
-1.0 0.25212
-0.8 0.21296
-0.6 0.17882
-0.4 0.14702
-0.2 0.11472
-0.0 0.08546
};
\addlegendentry{ $K=32$ RM 1 $C_2=4$ $L=8$}

\addplot[
    color=red,
    mark=o,
    thick,
    mark size=3,
]
table {
-2.0 0.42418
-1.8 0.39076
-1.6 0.35496
-1.4 0.31904
-1.2 0.28382
-1.0 0.24668
-0.8 0.2114
-0.6 0.1762
-0.4 0.14418
-0.2 0.11218
-0.0 0.08508
};
\addlegendentry{ $K=32$ RM 1 $C_2=7$ $L=2$}

\addplot[
    color=red,
    mark=+,
    thick,
    mark size=3,
]
table {
-2.0 0.4038
-1.8 0.37258
-1.6 0.33904
-1.4 0.30938
-1.2 0.27042
-1.0 0.23168
-0.8 0.19318
-0.6 0.15622
-0.4 0.12684
-0.2 0.09554
-0.0 0.07188
};
\addlegendentry{ $K=32$ RM 1 $C_2=7$ $L=8$}

\addplot[
    color=yellow,
    mark=square,
    thick,
    mark size=3,
]
table {
-1.2 0.6907
-1.0 0.64432
-0.8 0.6026
-0.6 0.56822
-0.4 0.53716
-0.2 0.5052
-0.0 0.48574
};
\addlegendentry{ $K=57$ RM 1 $C_2=4$ $L=2$}

\addplot[
    color=yellow,
    mark=star,
    thick,
    mark size=3,
]
table {
-1.2 0.657
-1.0 0.61568
-0.8 0.57174
-0.6 0.53766
-0.4 0.51318
-0.2 0.48438
-0.0 0.47516
};
\addlegendentry{ $K=57$ RM 1 $C_2=4$ $L=8$}

\addplot[
    color=yellow,
    mark=o,
    thick,
    mark size=3,
]
table {
-1.2 0.66502
-1.0 0.62118
-0.8 0.58256
-0.6 0.54974
-0.4 0.52644
-0.2 0.50114
-0.0 0.48462
};
\addlegendentry{ $K=57$ RM 1 $C_2=7$ $L=2$}

\addplot[
    color=yellow,
    mark=+,
    thick,
    mark size=3,
]
table {
-1.2 0.60766
-1.0 0.5653
-0.8 0.535
-0.6 0.50776
-0.4 0.49852
-0.2 0.47896
-0.0 0.46916
};
\addlegendentry{ $K=57$ RM 1 $C_2=7$ $L=8$}

\addplot[
    color=yellow,
    mark=square,
    thick,
    dashed,
    mark size=3,
]
table {
-1.2 0.6879
-1.0 0.6413
-0.8 0.59854
-0.6 0.55314
-0.4 0.52256
-0.2 0.49054
-0.0 0.47064
};
\addlegendentry{ $K=57$ RM 2 $C_2=4$ $L=2$}

\addplot[
    color=yellow,
    mark=star,
    thick,
    dashed,
    mark size=3,
]
table {
-1.2 0.65506
-1.0 0.60784
-0.8 0.56412
-0.6 0.5259
-0.4 0.50394
-0.2 0.4729
-0.0 0.45688
};
\addlegendentry{ $K=57$ RM 2 $C_2=4$ $L=8$}

\addplot[
    color=yellow,
    mark=o,
    thick,
    dashed,
    mark size=3,
]
table {
-1.2 0.66266
-1.0 0.61392
-0.8 0.57422
-0.6 0.53754
-0.4 0.51502
-0.2 0.48368
-0.0 0.46778
};
\addlegendentry{ $K=57$ RM 2 $C_2=7$ $L=2$}

\addplot[
    color=yellow,
    mark=+,
    thick,
    dashed,
    mark size=3,
]
table {
-1.2 0.6042
-1.0 0.55866
-0.8 0.52616
-0.6 0.49974
-0.4 0.48674
-0.2 0.4654
-0.0 0.45402
};
\addlegendentry{ $K=57$ RM 2 $C_2=7$ $L=8$}

\end{axis}
\end{tikzpicture}
  \caption{Missed detection ratios after SCL decoding, for transmissions including $C_1/2$ $N_1=128$ and $C_1/2$ $N_2=256$.}
  \label{fig:MD_SCL}
\end{figure*}


We have built a simulation environment to evaluate the feasibility to use polar codes in a blind detection framework.
We have performed simulations to evaluate the BLER, MDR and FAR of the proposed blind detection scheme under a variety of parameters. Three block lengths ($128$, $256$, $512$) and four information lengths ($8$, $16$, $32$, $57$) have been considered. The position of the RNTI has been selected according to two operation modes:
\begin{itemize}
 \item RNTI Mode 1 (RM1): RNTI bits are the most reliable after the $K$ information bits. 
 \item RNTI Mode 2 (RM2): RNTI bits are the most reliable, while the $K$ information bits are the most reliable after the RNTI bits.
\end{itemize}
Moreover, four SCL candidates $C_2$ ($4$, $5$, $6$, $7$), and three list sizes $L$ ($2$, $4$, $8$), have been considered as well.

Fig.~\ref{fig:BLER} depicts the BLER of the simulated codes after SC decoding only. It can be seen that the difference between RM1 and RM2 is generally negligible. A missed detection occurs when the UE fails to identify its RNTI among the received frames.
Fig.~\ref{fig:MD_SCL} depicts the MDR after SCL decoding, where MDR is defined as the number of missed detections over the number of transmissions in which the UE RNTI was sent. MDR simulations consider $C_1/2$ candidates of length $N_1$, and $C_1/2$ candidates of length $N_2$, all with an information length of $K_1 = K_2 = K$ bits. The UE RNTI is randomly transmitted through one of the $C_1$ possible codes. 
The drawn curves consider the extreme values of the $C_2$ and $L$ simulation space, i.e $C_2=(4,7)$ and $L=(2,8)$. Performance of the intermediate values sits in between the portrayed ones.
It can be observed that increasing $C_2$ and $L$ leads to better MDR, regardless of the code lengths and rates. Increasing $C_2$ rises the probability of having, among the $C_2$ SCL candidates, the one whose RNTI matches the UE RNTI. A larger $L$ improves the error correction-performance of the SCL algorithm.
RM2 has a substantial advantage over RM1 when MDR is high, and grants slight improvements at lower MDR, as shown on the yellow and blue curves in Fig.~\ref{fig:MD_SCL}.
In general, the MDR curve is shown to be substantially lower than the BLER curve.

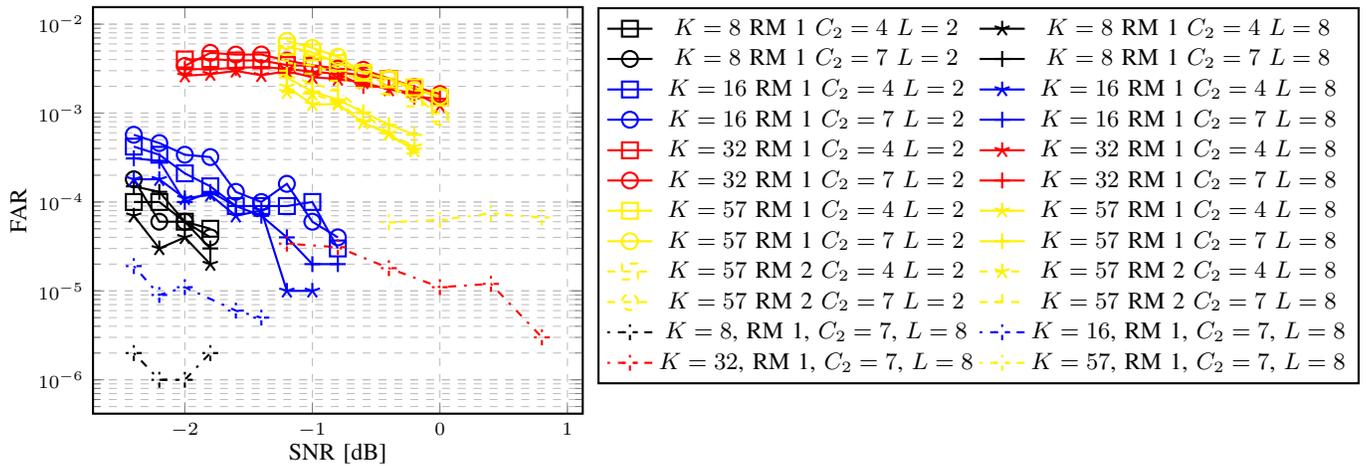
\begin{figure*}[t!]
  \centering
  \begin{tikzpicture}
  \pgfplotsset{
    label style = {font=\fontsize{9pt}{7.2}\selectfont},
    tick label style = {font=\fontsize{7pt}{7.2}\selectfont},
  }
\small

\begin{axis}[
	scale = 0.95,
    ymode=log,
    xlabel={SNR [dB]}, xlabel style={yshift=0.8em},
    ylabel={FAR}, ylabel style={yshift=-0.75em},%
    grid=both,
    ymajorgrids=true,
    xmajorgrids=true,
    grid style=dashed,
    thick,
    mark size=3,
    legend columns=2,
    legend pos=outer north east,
]

\addplot[
    color=black,
    mark=square,
    thick,
    mark size=3,
]
table {
-2.4 0.00010
-2.2 0.00010
-2.0 0.00006
-1.8 0.00005
};
\addlegendentry{ $K=8$ RM 1 $C_2=4$ $L=2$}

\addplot[
    color=black,
    mark=star,
    thick,
    mark size=3,
]
table {
-2.4 0.00007
-2.2 0.00003
-2.0 0.00004
-1.8 0.00002
};
\addlegendentry{ $K=8$ RM 1 $C_2=4$ $L=8$}

\addplot[
    color=black,
    mark=o,
    thick,
    mark size=3,
]
table {
-2.4 0.00018
-2.2 0.00006
-2.0 0.00006
-1.8 0.00004
};
\addlegendentry{ $K=8$ RM 1 $C_2=7$ $L=2$}

\addplot[
    color=black,
    mark=+,
    thick,
    mark size=3,
]
table {
-2.4 0.00015
-2.2 0.00013
-2.0 0.00006
-1.8 0.00003
};
\addlegendentry{ $K=8$ RM 1 $C_2=7$ $L=8$}

\addplot[
    color=blue,
    mark=square,
    thick,
    mark size=3,
]
table {
-2.4 0.00042
-2.2 0.00034
-2.0 0.00021
-1.8 0.00015
-1.6 0.00009
-1.4 0.00009
-1.2 0.00009
-1.0 0.00010
-0.8 0.00003
};
\addlegendentry{ $K=16$ RM 1 $C_2=4$ $L=2$}

\addplot[
    color=blue,
    mark=star,
    thick,
    mark size=3,
]
table {
-2.4 0.00018
-2.2 0.00018
-2.0 0.00011
-1.8 0.00012
-1.6 0.00007
-1.4 0.00008
-1.2 0.00001
-1.0 0.00001
-0.8 0
};
\addlegendentry{ $K=16$ RM 1 $C_2=4$ $L=8$}

\addplot[
    color=blue,
    mark=o,
    thick,
    mark size=3,
]
table {
-2.4 0.00057
-2.2 0.00046
-2.0 0.00034
-1.8 0.00032
-1.6 0.00013
-1.4 0.00010
-1.2 0.00016
-1.0 0.00006
-0.8 0.00004
};
\addlegendentry{ $K=16$ RM 1 $C_2=7$ $L=2$}

\addplot[
    color=blue,
    mark=+,
    thick,
    mark size=3,
]
table {
-2.4 0.00031
-2.2 0.00029
-2.0 0.00010
-1.8 0.00013
-1.6 0.00009
-1.4 0.00007
-1.2 0.00004
-1.0 0.00002
-0.8 0.00002
};
\addlegendentry{ $K=16$ RM 1 $C_2=7$ $L=8$}

\addplot[
    color=red,
    mark=square,
    thick,
    mark size=3,
]
table {
-2.0 0.00401
-1.8 0.00405
-1.6 0.00391
-1.4 0.00391
-1.2 0.00347
-1.0 0.00338
-0.8 0.00289
-0.6 0.00264
-0.4 0.00241
-0.2 0.00191
-0.0 0.00149
};
\addlegendentry{ $K=32$ RM 1 $C_2=4$ $L=2$}

\addplot[
    color=red,
    mark=star,
    thick,
    mark size=3,
]
table {
-2.0 0.00263
-1.8 0.00274
-1.6 0.00298
-1.4 0.00266
-1.2 0.00296
-1.0 0.00253
-0.8 0.00242
-0.6 0.00221
-0.4 0.00184
-0.2 0.00169
-0.0 0.00121
};
\addlegendentry{ $K=32$ RM 1 $C_2=4$ $L=8$}

\addplot[
    color=red,
    mark=o,
    thick,
    mark size=3,
]
table {
-2.0 0.00339
-1.8 0.00476
-1.6 0.00455
-1.4 0.00456
-1.2 0.00389
-1.0 0.00346
-0.8 0.00316
-0.6 0.00307
-0.4 0.00245
-0.2 0.00197
-0.0 0.00164
};
\addlegendentry{ $K=32$ RM 1 $C_2=7$ $L=2$}

\addplot[
    color=red,
    mark=+,
    thick,
    mark size=3,
]
table {
-2.0 0.00313
-1.8 0.00311
-1.6 0.00323
-1.4 0.00329
-1.2 0.00315
-1.0 0.00292
-0.8 0.00276
-0.6 0.00202
-0.4 0.00194
-0.2 0.00153
-0.0 0.00145
};
\addlegendentry{ $K=32$ RM 1 $C_2=7$ $L=8$}

\addplot[
    color=yellow,
    mark=square,
    thick,
    mark size=3,
]
table {
-1.2 0.00562
-1.0 0.00481
-0.8 0.00365
-0.6 0.00284
-0.4 0.00239
-0.2 0.00195
-0.0 0.00149
};
\addlegendentry{ $K=57$ RM 1 $C_2=4$ $L=2$}

\addplot[
    color=yellow,
    mark=star,
    thick,
    mark size=3,
]
table {
-1.2 0.00198
-1.0 0.00125
-0.8 0.00126
-0.6 0.00078
-0.4 0.00058
-0.2 0.00038
};
\addlegendentry{ $K=57$ RM 1 $C_2=4$ $L=8$}

\addplot[
    color=yellow,
    mark=o,
    thick,
    mark size=3,
]
table {
-1.2 0.00647
-1.0 0.00542
-0.8 0.00431
-0.6 0.00283
-0.4 0.00244
-0.2 0.00180
-0.0 0.00155
};
\addlegendentry{ $K=57$ RM 1 $C_2=7$ $L=2$}

\addplot[
    color=yellow,
    mark=+,
    thick,
    mark size=3,
]
table {
-1.2 0.00251
-1.0 0.00175
-0.8 0.00142
-0.6 0.00102
-0.4 0.00074
-0.2 0.00057
};
\addlegendentry{ $K=57$ RM 1 $C_2=7$ $L=8$}

\addplot[
    color=yellow,
    mark=square,
    thick,
    dashed,
    mark size=3,
]
table {
-1.2 0.00368
-1.0 0.00341
-0.8 0.00284
-0.6 0.00237
-0.4 0.00201
-0.2 0.00146
-0.0 0.00099
};
\addlegendentry{ $K=57$ RM 2 $C_2=4$ $L=2$}

\addplot[
    color=yellow,
    mark=star,
    thick,
    dashed,
    mark size=3,
]
table {
-1.2 0.00172
-1.0 0.00164
-0.8 0.00132
-0.6 0.00080
-0.4 0.00059
-0.2 0.00042
};
\addlegendentry{ $K=57$ RM 2 $C_2=4$ $L=8$}

\addplot[
    color=yellow,
    mark=o,
    thick,
    dashed,
    mark size=3,
]
table {
-1.2 0.00476
-1.0 0.00423
-0.8 0.00399
-0.6 0.00296
-0.4 0.00198
-0.2 0.00164
-0.0 0.00090
};
\addlegendentry{ $K=57$ RM 2 $C_2=7$ $L=2$}

\addplot[
    color=yellow,
    mark=+,
    thick,
    dashed,
    mark size=3,
]
table {
-1.2 0.00269
-1.0 0.00215
-0.8 0.00180
-0.6 0.00100
-0.4 0.00059
-0.2 0.00035
};
\addlegendentry{ $K=57$ RM 2 $C_2=7$ $L=8$}

\addplot[
    color=black,
    mark=+,
    dash pattern=on 1pt off 3pt on 3pt off 3pt,
    thick,
    mark size=3,
]
table {
-2.4 0.000002
-2.2 0.000001
-2.0 0.000001
-1.8 0.000002
};
\addlegendentry{ $K=8$, RM 1, $C_2=7$, $L=8$}

\addplot[
    color=blue,
    mark=+,
    thick,
    dash pattern=on 1pt off 3pt on 3pt off 3pt,
    mark size=3,
]
table {
-2.4 0.000019
-2.2 0.000009
-2.0 0.000011 
-1.8 0.000006
-1.6 0.000006
-1.4 0.000005
};
\addlegendentry{ $K=16$, RM 1, $C_2=7$, $L=8$}

\addplot[
    color=red,
    mark=+,
    thick,
    dash pattern=on 1pt off 3pt on 3pt off 3pt,
    mark size=3,
]
table {
-1.2 0.000034
-0.8 0.000031
-0.4 0.000018
0.0 0.000011
0.4 0.000012
0.8 0.000003
};
\addlegendentry{ $K=32$, RM 1, $C_2=7$, $L=8$}

\addplot[
    color=yellow,
    mark=+,
    thick,
    dash pattern=on 1pt off 3pt on 3pt off 3pt,
    mark size=3,
]
table {
-0.4 0.000059
-0.0 0.000063
0.4 0.000074
0.8 0.000067
};
\addlegendentry{ $K=57$, RM 1, $C_2=7$, $L=8$}

\end{axis}
\end{tikzpicture}
  \caption{False alarm ratios after SCL decoding, for transmissions including $C_1/2$ $N_1=128$ and $C_1/2$ $N_2=256$.}
  \label{fig:FA_SCL}
\end{figure*}

The false alarm curves shown in Fig.~\ref{fig:FA_SCL} report the combination of Type-1 and Type-2 errors. All curves have been obtained over $10^5$ transmissions, in half of which the UE RNTI was sent. The blue and black curves have been obtained with very few counted errors ($<10$): given the total number of simulated transmissions, we can reliably upper bound the FAR at $<10^{-3}$. 
These curves, however, have been obtained with \emph{sequential} RNTIs, as a worst case: even if the RNTI is $16$-bit long, the $C_1$ RNTIs go from $0$ to $C_1-1$. This increases the probability of false alarms. 
Thus, the last four curves in Fig.~\ref{fig:FA_SCL} (dash-dot curve pattern) show results obtained with the RNTIs of the $C_1$ candidates assuming random values over the full $16$-bit dynamic; these curves show orders of magnitude lower FAR with respect to the sequential RNTIs case. They have been obtained over $10^6$ transmissions. For $K=8$ and $K=16$, $<10$ errors have been counted over the $10^6$ trials, thus we can reliably upper bound the FAR at $10^{-4}$. Simulations show that the MDR is not affected by the randomization of the RNTI values.

\section{Early Stopping} \label{sec:earlystop}


This section presents an early stopping criterion effective in reducing the average time needed by the second phase of the blind detection scheme, the SCL decoding.
The first phase of the proposed blind detection scheme requires the full decoding of each candidate, in order to identify the $C_2$ codewords that will be decoded with SCL. In the SCL decoding phase, however, all codewords whose RNTI does not match the UE RNTI will surely be discarded. Thus, as soon as the RNTI is shown to be different, the decoding can be interrupted. Since SC-based decoding algorithms estimate codeword bits sequentially, the RNTI evaluation can be performed every time an RNTI bit is estimated. In case the estimated bit is different from the UE RNTI bit, the decoding is stopped. 
Algorithm~\ref{alg:earlystop} describes in detail the proposed early stopping criterion. Let us consider the set of $L$ newly estimated bits $E$. If $E$ corresponds to an RNTI bit, each path $j$, with $0\le j < L$, compares $E_j$ to the related UE RNTI bit. It they are not equal, path $j$ is deactivated, and when all paths are deactivated, the SCL decoder is stopped.

		\begin{algorithm}[t!] 
	   \SetKwInOut{Input}{input} \SetKwInOut{Output}{output}
	   \DontPrintSemicolon

	   \Input{Set of the $L$ estimated bits $E$}
	   \Input{Next RNTI bit index $w$}
	   \Output{Next RNTI bit index $w$}
	   \Begin{
	      \If{$E\in\text{RNTI}$}{
		\For{$j=0:L-1$}{
		   \If{$E_j==\text{RNTI}_{w}$}{
		      Path $j$ is maintained active
		    }
		    \Else{
		      Path $j$ is deactivated
		    }
		}
		$w=w+1$
	      }
	      \Else{
		  All paths are maintained active
	      }
	      \If{all $j$ paths are deactivated}{
		Stop SCL decoder
	      }
	   }
	   \caption{SCL early stopping criterion}
	   \label{alg:earlystop}
\end{algorithm}

\begin{figure*}[t!]
  \centering
  \begin{tikzpicture}
  \pgfplotsset{
    label style = {font=\fontsize{9pt}{7.2}\selectfont},
    tick label style = {font=\fontsize{7pt}{7.2}\selectfont},
  }

\begin{axis}[
	scale = 0.95,
    xlabel={SNR [dB]}, xlabel style={yshift=0.8em},
    ylabel={Average estimated bit \%}, ylabel style={yshift=-0.75em},%
    grid=both,
    ymajorgrids=true,
    xmajorgrids=true,
    grid style=dashed,
    thick,
    mark size=3,
    legend columns=2,
    legend pos=outer north east,
]

\addplot[
    color=black,
    mark=square,
    thick,
    mark size=3,
]
table {
-2.4 60.4
-2.2 60.7
-2.0 60.7
-1.8 60.9
};
\addlegendentry{$N=128$ $K=8$ }

\addplot[
    color=black,
    mark=square,
    thick,
    dashed,
    mark size=3,
]
table {
-2.4 50.8
-2.2 50.8
-2.0 50.8
-1.8 50.8
};
\addlegendentry{$N=128$ $K=8$ - no UE RNTI}

\addplot[
    color=blue,
    mark=square,
    thick,
    mark size=3,
]
table {
-2.4 41.9
-2.2 42.2
-2.0 42.3
-1.8 42.3
-1.6 42.7
-1.4 43.2
};
\addlegendentry{$N=128$ $K=16$ }

\addplot[
    color=blue,
    mark=square,
    thick,
    dashed,
    mark size=3,
]
table {
-2.4 32.6
-2.2 32.6
-2.0 32.6
-1.8 32.6
-1.6 32.6
-1.4 32.6
};
\addlegendentry{$N=128$ $K=16$ - no UE RNTI}

\addplot[
    color=red,
    mark=square,
    thick,
    mark size=3,
]
table {
-2.0 39.4
-1.8 39.5
-1.6 39.9
-1.4 40.2
-1.2 40.5
-1.0 41.1
-0.8 41.7
-0.6 42.4
-0.4 43.0
-0.2 43.5
0.0 44.3
0.2 44.7
0.4 45.2
0.6 45.6
};
\addlegendentry{$N=128$ $K=32$ }

\addplot[
    color=red,
    mark=square,
    thick,
    dashed,
    mark size=3,
]
table {
-2.0 32.3
-1.8 32.3
-1.6 32.3
-1.4 32.3
-1.2 32.3
-1.0 32.3
-0.8 32.3
-0.6 32.3
-0.4 32.3
-0.2 32.3
0.0 32.3
0.2 32.3
0.4 32.3
0.6 32.3
};
\addlegendentry{$N=128$ $K=32$ - no UE RNTI}

\addplot[
    color=yellow,
    mark=square,
    thick,
    mark size=3,
]
table {
-0.4 25.3
-0.2 25.3
0.0 25.4
0.2 25.5
0.4 26.0
0.6 26.1
0.8 27.3
1.0 27.7
1.2 28.3
1.4 29.4
1.6 30.2
};
\addlegendentry{$N=128$ $K=57$ }

\addplot[
    color=yellow,
    mark=square,
    thick,
    dashed,
    mark size=3,
]
table {
-0.4 20.3
-0.2 20.3
0.0 20.3
0.2 20.3
0.4 20.3
0.6 20.3
0.8 20.3
1.0 20.3
1.2 20.3
1.4 20.3
1.6 20.3
};
\addlegendentry{$N=128$ $K=57$ - no UE RNTI}

\addplot[
    color=black,
    mark=+,
    thick,
    mark size=3,
]
table {
-2.4 67.7
-2.2 67.9
-2.0 68.2
-1.8 68.6
};
\addlegendentry{$N=256$ $K=8$ }

\addplot[
    color=black,
    mark=+,
    thick,
    dashed,
    mark size=3,
]
table {
-2.4 51.6
-2.2 51.6
-2.0 51.6
-1.8 51.6
};
\addlegendentry{$N=256$ $K=8$ - no UE RNTI}

\addplot[
    color=blue,
    mark=+,
    thick,
    mark size=3,
]
table {
-2.4 58.8
-2.2 61.2
-2.0 64.4
-1.8 66.6
-1.6 67.9
-1.4 68.3
};
\addlegendentry{$N=256$ $K=16$ }

\addplot[
    color=blue,
    mark=+,
    thick,
    dashed,
    mark size=3,
]
table {
-2.4 41.4
-2.2 41.4
-2.0 41.4
-1.8 41.4
-1.6 41.4
-1.4 41.4
};
\addlegendentry{$N=256$ $K=16$ - no UE RNTI}

\addplot[
    color=red,
    mark=+,
    thick,
    mark size=3,
]
table {
-2.0 45.7
-1.8 47.4
-1.6 49.1
-1.4 51.1
-1.2 53.2
-1.0 55.0
-0.8 56.2
-0.6 56.7
-0.4 56.5
-0.2 56.6
0.0 56.2
0.2 56.2
0.4 56.2
0.6 56.2
0.8 56.2
};
\addlegendentry{$N=256$ $K=32$ }

\addplot[
    color=red,
    mark=+,
    thick,
    dashed,
    mark size=3,
]
table {
-2.0 32.4
-1.8 32.4
-1.6 32.4
-1.4 32.4
-1.2 32.4
-1.0 32.4
-0.8 32.4
-0.6 32.4
-0.4 32.4
-0.2 32.4
0.0 32.4
0.2 32.4
0.4 32.4
0.6 32.4
};
\addlegendentry{$N=256$ $K=32$ - no UE RNTI}

\addplot[
    color=yellow,
    mark=+,
    thick,
    mark size=3,
]
table {
-0.4 43.1
-0.2 44.5
0.0 45.8
0.2 47.4
0.4 49.1
0.6 50.8
0.8 52.4
1.0 54.5
1.2 56.2
1.4 57.1
1.6 58.1
};
\addlegendentry{$N=256$ $K=57$ }

\addplot[
    color=yellow,
    mark=+,
    thick,
    dashed,
    mark size=3,
]
table {
-0.4 32.1
-0.2 32.1
0.0 32.1
0.2 32.1
0.4 32.1
0.6 32.1
0.8 32.1
1.0 32.1
1.2 32.1
1.4 32.1
1.6 32.1
};
\addlegendentry{$N=256$ $K=57$ - no UE RNTI}

\end{axis}
\end{tikzpicture}
  \caption{Average percentage of SCL estimated bits with early stopping, randomized transmitted RNTIs, RM3.}
  \label{fig:avg_est_RM3}
\end{figure*}
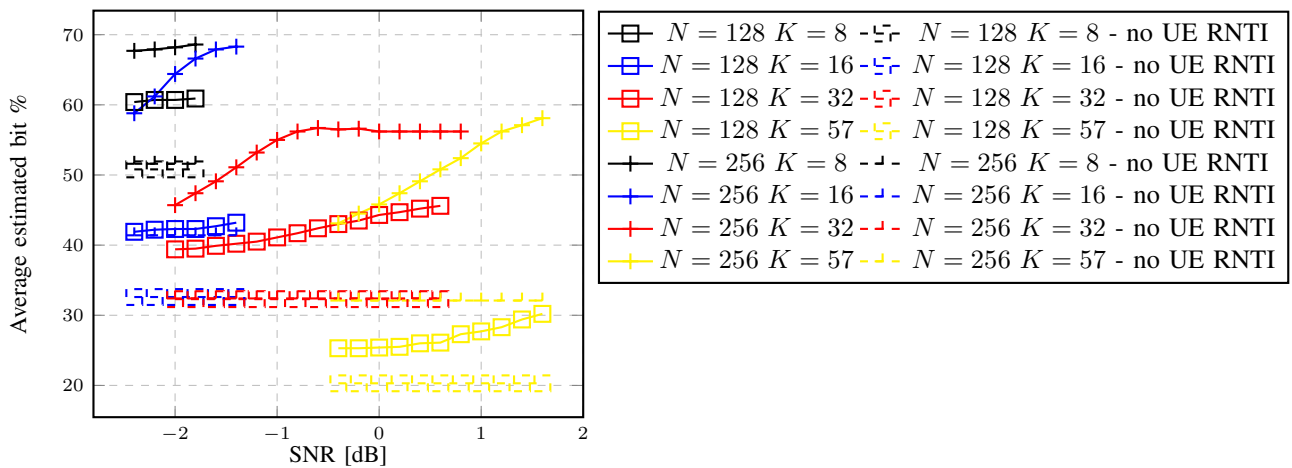

The average number of estimated bits is heavily dependent on the position of the RNTI bits within the polar code. In particular, if the bits assigned to the RNTI are towards the left of the decoding tree, a non-matching RNTI will be identified earlier in the decoding process, leading to a lower average number of estimated bits. We consequently evaluated the performance of the proposed early stopping criterion when the RNTI bits are assigned to the leftmost positions among the $K+16$ most reliable ones: we call this RNTI bit selection method RNTI Mode 3 (RM3). RM3 selects bits of intermediate reliability between RM1 and RM2: since the difference in terms of error correction performance between RM1 and RM2 is negligible, RM3 does not cause any BER/FER degradation.
Fig.~\ref{fig:avg_est_RM3} shows the average percentage of estimated bits when the proposed early stopping criterion is applied, together with RM3 and RNTI randomization. These results consider each of the $C_2$ candidates separately, since the number of candidates of length $N_1$ and $N_2$ decoded with SCL depends on the SC path metrics, and thus on channel noise. Moreover, we have observed that the average number of estimated bits undergoes negligible variations when different list sizes $L$ are considered: nevertheless, the reported curves are averaged between $L=2$ and $L=8$. The solid curves have been obtained with the UE RNTI being sent through the considered code. It can be seen that as the channel conditions improve, the number of estimated bits increases until a plateau region is reached. This is due to the fact that when the SNR is low, it is more likely that the codeword with the UE RNTI will not be among the $C_2$ SCL candidates. Thus, even if there are errors in the codeword, the SCL decoders will easily encounter RNTI bits different from the UE RNTI early in the decoding process. As the SNR increases, the codeword with the UE RNTI will be among the $C_2$ candidates with rising probability. In parallel, the SCL decoder to which it is assigned will not interrupt the decoding, leading to $100\%$ estimated bits, while the other $C_2-1$ decoders will stop the decoding early, finally settling the average estimated bit percentage at a stable value. This is easily noticed in the $N=256$, $K=32$ red curve, where from SNR=$-1$dB onwards the percentage stays at $56.2\%$. This percentage would be higher if RM1 was used ($81.5\%$ in this case). The dashed curves have been obtained simulating cases in which the UE RNTI was not sent. It is possible to see how the average estimated bit percentage remains constant as the SNR changes: since among the $C_2$ candidates there is never one whose RNTI matches the UE RNTI, all SCL decoders tend to stop the decoding early.

\section{Detection Speed} \label{sec:decspeed}

\begin{table}[t!]
	\centering
	\caption{Time-Steps Requirements}
	\label{tab:timestep}
		\setlength{\extrarowheight}{1.7pt}
\footnotesize
	\begin{tabular}{c|c|c|c|c|c|}
	\cline{2-6}
	& \multicolumn{5}{c|}{Decoding Algorithm} \\
	\cline{2-6}
	& SC & Fast-SSC & SCL & SSCL & {\scriptsize Fast-SSCL~$L=2$}\\
	\hline
	\multicolumn{1}{|c|}{$\mathcal{P}(128,8)$} & $254$ & $43$ & $278$ & $79$ & $71$\\
	\hline
	\multicolumn{1}{|c|}{$\mathcal{P}(128,16)$} & $254$ & $46$ & $286$ & $81$ & $67$\\
	\hline
	\multicolumn{1}{|c|}{$\mathcal{P}(128,32)$} & $254$ & $49$ & $302$ & $112$ & $86$\\
	\hline
	\multicolumn{1}{|c|}{$\mathcal{P}(128,57)$} & $254$ & $52$ & $327$ & $134$ & $84$\\
	\hline
	\multicolumn{1}{|c|}{$\mathcal{P}(256,8)$} & $510$ & $94$ & $534$ & $122$ & $120$\\
	\hline
	\multicolumn{1}{|c|}{$\mathcal{P}(256,16)$} & $510$ & $97$ & $542$ & $130$ & $125$\\
	\hline
	\multicolumn{1}{|c|}{$\mathcal{P}(256,32)$} & $510$ & $109$ & $558$ & $163$ & $149$\\
	\hline
	\multicolumn{1}{|c|}{$\mathcal{P}(256,57)$} & $510$ & $127$ & $583$ & $226$ & $203$\\
	\hline
	\multicolumn{1}{|c|}{$\mathcal{P}(512,8)$} & $1022$ & $37$ & $1046$ & $71$ & $64$\\
	\hline
	\multicolumn{1}{|c|}{$\mathcal{P}(512,16)$} & $1022$ & $64$ & $1054$ & $110$ & $101$\\
	\hline
	\multicolumn{1}{|c|}{$\mathcal{P}(512,32)$} & $1022$ & $85$ & $1070$ & $140$ & $124$\\
	\hline
	\multicolumn{1}{|c|}{$\mathcal{P}(512,57)$} & $1022$ & $91$ & $1095$ & $193$ & $163$\\
	\hline
	\end{tabular}

\end{table}

The blind detection process in LTE needs to be performed in $16\mu$s: however, ongoing discussions in the 5G standardization process might shorten the available time to $4\mu$s. We thus analyze the duration of the blind detection process based on polar codes, according to the system parameters. The analysis does not take in account the early stopping criterion, thus providing worst-case results. The average latency gain brought by early stopping of SCL is dependent on the code size of the $C_2$ candidates, that cannot be foreseen at design time.

Assuming to decode with SC all the $N_1$-length locations first, and the $N_2$-length locations after, the number of time-steps required to complete the different phases is the following:
\begin{equation} \label{eq:lat2}
 T_{\text{bd}}=\left\lceil \frac{C_1}{N_{\text{SC}}} \right\rceil \left(\frac{T^1_{\text{SC}}}{2}+\frac{T^2_{\text{SC}}}{2}\right) + T_{\text{sort}} + \left\lceil \frac{C_2}{N_{\text{SCL}}} \right\rceil T_{\text{SCL}}
\end{equation}
where $N_{\text{SC}}$ and $N_{\text{SCL}}$ are the number of SC and SCL decoders working in parallel, and $T^1_{\text{SC}}$ and $T^2_{\text{SC}}$ are the SC decoding latencies for codes of length $N_1$ and $N_2$, respectively. $T_{\text{SCL}}$ is the decoding latency of an SCL decoder, while $T_{\text{sort}}$ is the number of time steps required to obtain the $C_2$ SCL candidates out of the $C_1$ candidate locations through sorting.
The worst case for $T_{\text{SC}}$ and $T_{\text{SCL}}$ occurs when the standard SC and SCL algorithms are applied, without exploiting tree-pruning techniques that rely on constituent codes, like in Fast-SSC \cite{sarkis}, SSCL \cite{hashemi_SSCL} and Fast-SSCL \cite{hashemi_FSSCL}. In the traditional SC and SCL cases, the decoding latencies can be expressed as:
\begin{equation*}
T^i_{\text{SC}}=2N_i-2
\end{equation*}
\begin{equation*}
 T_{\text{SCL}}=\max(2N_1+K_1,2N_2+K_2)+RNTI_b-2
\end{equation*}
where $RNTI_b$ represents the number of bits assigned to the RNTI. In our case, we can fix parameters $C_1=44$ and $RNTI_b=16$, and estimate $T_{\text{sort}}$, whose contribution to the latency is minimal, as $T_{\text{sort}}=C_2$. 
The worst case sees $N_1=512$, $N_2=256$, $K_1=K_2=57$. The $4\mu$s mark is achieved with $f=800$~MHz, $N_{\text{SC}}=22$, $N_{\text{SCL}}=C_2$.

Considering the Fast-SSC, SSCL and Fast-SSCL algorithms allows to exploit particular patterns of frozen and information bits to reduce the decoding latency and thus the complexity needed to reach the $4\mu$s target. The achievable gain depends on the code structure. 
In our case, the number of time steps necessary for the decoding of each considered code with different decoding algorithms is detailed in Table~\ref{tab:timestep}. The worst case occurs for $N_1=N_2=256$, $K_1=57$, $K_2=32$. Results are valid for RM1, RM2, and RM3, since for the decoding process the RNTI bits are considered information bits.

Table~\ref{tab:latency} reports combinations of parameters that satisfy the $4\mu$s target. 
It is possible to see that the faster decoding process of Fast-SSC and SSCL allows to drastically reduce the resource needed to meet the latency target with respect to standard SC and SCL.



\begin{table}[t!]
	\centering
	\caption{Parameters needed to meet the $4\mu$s target}
	\label{tab:latency}
		\setlength{\extrarowheight}{1.7pt}
	\begin{tabular}{|c|c|c|c|c|}
	\hline
	 Algorithm & $f$ [MHz] & $N_{\text{SC}}$ & $N_{\text{SCL}}$ & Latency [$\mu$s] \\
	 \hline
	 \hline
	 SC + SCL & $800$ & $22$ & $C_2$ & $3.9$ \\
	 \hline
	  & $300$ & $11$ & $C_2/2$ & $3.8$ \\
	 \multirow{2}{*}{Fast-SSC +}& $400$ & $8$ & $C_2/2$ & $3.7$ \\
	 \multirow{2}{*}{SSCL}& $500$ & $5$ & $C_2/2$ & $3.5$ \\
	 & $600$ & $4$ & $C_2/2$ & $3.8$ \\
	 & $700$ & $3$ & $C_2/2$ & $4.0$ \\
	 \hline
	 & $300$ & $11$ & $C_2/2$ & $3.7$ \\
	 \multirow{2}{*}{Fast-SSC +}& $400$ & $7$ & $C_2/2$ & $3.9$ \\
	 \multirow{2}{*}{Fast-SSCL, $L=2$}& $500$ & $5$ & $C_2/2$ & $3.4$ \\
	 & $600$ & $4$ & $C_2/2$ & $3.8$ \\
	 & $700$ & $3$ & $C_2/2$ & $3.9$ \\
	 \hline
	\end{tabular}

\end{table}

\section{Conclusion} \label{sec:conc}

In this work we have proposed a novel blind detection scheme that relies on polar codes and does not need a cyclic redundancy check. Simulation results show that the scheme can easily outperform the 3GPP LTE/LTE-Advanced requirements in terms of missed-detection rate and false-alarm rate. An early stopping criterion is proposed and evaluated, showing that the average number of operations in the second phase of the blind detection scheme can be substantially reduced at no cost in performance. The time complexity of the blind detection scheme is then analyzed: using common polar code decoding algorithms, the $4\mu$s latency target can be met with a variety of system parameter combinations.


\end{document}